\documentclass[a4paper]{jpconf}
\usepackage{epsf}
\usepackage{amsmath,amsfonts,amssymb}
\usepackage{graphicx,color}

\def\thetable{\arabic{section}.\arabic{table}}

\newcommand{\nc}{\newcommand}
\nc{\rnc}{\renewcommand}
\nc{\nn}{\nonumber}
\nc{\db}{\displaybreak[0]\\}
\nc{\ds}{\displaystyle}
\rnc{\c}[1]{\textcolor{blue}{#1}}
\rnc{\[}{\left[}
\rnc{\]}{\right]}
\rnc{\(}{\left(}
\rnc{\)}{\right)}
\nc{\lt}{\left\{}
\nc{\rt}{\right\}}
\rnc{\o}{\omega}
\rnc{\a}{\alpha}
\rnc{\b}{\beta}
\nc{\lam}{\lambda}
\nc{\sig}{\sigma}
\nc{\eps}{\epsilon}
\nc{\vp}{\varphi}
\nc{\del}{\delta}
\nc{\D}{\Delta}
\rnc{\th}{\theta}
\nc{\Th}{\Theta}
\nc{\z}{\zeta}
\nc{\g}{\gamma}
\nc{\ch}{\cosh}
\nc{\sh}{\sinh}
\nc{\bra}{\langle}
\nc{\ket}{\rangle}
\rnc{\i}{{\rm i}}
\rnc{\d}{{\rm d}}
\nc{\Res}[2]{{\rm Res}\[\left.#1\right|_{#2}\]}
\nc{\intall}{\int_{-\infty}^\infty}
\nc{\inth}{\int_0^{\infty}}
\nc{\atan}{\tan^{-1}}
\nc{\sign}{\mbox{sign}}
\nc{\xxz}{H_{\text{XXZ}}}
\nc{\sm}[2]{\sum_{#1}^{#2}}
\nc{\til}{\tilde}
\nc{\hh}{\check{h}}
\nc{\RR}{\check{R}}
\nc{\dlam}{\frac d{d\lam}}
\nc{\cd}{\cdots}
\nc{\re}{\eqref}
\nc{\fr}[2]{\frac{#1}{#2}}
\nc{\obib}[2]{\frac{d {#1}}{d {#2}}}
\nc{\pbib}[2]{\frac{\partial {#1}}{\partial {#2}}}
\rnc{\k}{\kappa}

\begin{document}

\title{
Crossover temperature of  the spin-1/2 XXZ chain with an impurity
}

\author{Ryoko Yahagi$^1$, Jun Sato$^2$ and Tetsuo Deguchi$^3$}

\address{$^{1}$ Department of Physics, Graduate School of Humanities and Sciences, Ochanomizu University, 
2-1-1 Ohtsuka, Bunkyo-ku, Tokyo 112-8610, Japan}
\address{$^2$ Research Center for Advanced Science and Technology, University of Tokyo, \\
4-6-1 Komaba, Meguro-ku, Tokyo 153-8904, Japan}
\address{$^3$ Department of Physics, Faculty of Core Research, Ochanomizu University, \\
2-1-1 Ohtsuka, Bunkyo-ku, Tokyo 112-8610, Japan}

\ead{$^1$yahagi@hep.phys.ocha.ac.jp, 
$^2$jsato@jamology.rcast.u-tokyo.ac.jp, \\
$^3$deguchi@phys.ocha.ac.jp}

\begin{abstract}
We study exactly the effect of an impurity in the\ interacting quantum spin chain at low temperature by solving the integrable spin-1/2 XXZ periodic chain with an impurity through the algebraic and thermal Bethe ansatz methods. In particular, we investigate how the crossover temperature for the impurity specific heat depends on the impurity parameter, i.e. the coupling of the impurity to other spins,  and show that  it is consistent with the analytic expression that is obtained by setting the impurity susceptibility to be proportional to the inverse of the crossover temperature.  In the model, two types of crossover behavior appear: one from the high-temperature regime to the low-temperature Kondo regime and another from the $N$-site homogeneous chain to the $(N-1)$-site chain with a decoupled free impurity spin, with respect to the temperature and  the impurity parameter, respectively. 
\end{abstract}

\section{Introduction}

The Kondo effect was discovered experimentally in the 1930s \cite{H} and was theoretically explained first in the 1960s \cite{K64}. 
However, the Kondo problem and related subjects in many-body problem still evoke considerable interest in various theoretical and experimental researches \cite{Nozieres,Hewson,Kondo}.  
First, quantum impurity systems show universal critical behavior at low  temperature, 
which is characterized by the ratio of the impurity susceptibility $\chi_{\rm imp}$ 
to the impurity specific heat $c_{\rm imp}$ divided by temperature $T$: 
$r =  ({\pi^2}/{3})  {\chi_{\rm imp}}/(c_{\rm imp}/T)$  \cite{Wi}.   
We call it the Wilson ratio.  It was exactly shown by the Bethe ansatz  
that it is given by 2 for the Kondo model \cite{AFL,TW}.
Furthermore, the effect of an impurity embedded in a one-dimensional interacting quantum system 
or  in a Tomonaga-Luttinger liquid \cite{TL} has been one of the topics evoking considerable interest during the 1990s and during the last decade. 
It has been investigated by different methods such as renormalization group techniques \cite{Kane,Furusaki,Matveev}, conformal field theories (CFT) \cite{AL,Frojdh}, numerical techniques with CFT \cite{Eggert-Affleck,Oshikawa}, and 
the Bethe-ansatz method \cite{AJ,Schlottmann1991,Sacramento,Eckle,FZ1997,KZ,Sch,ZK2000,Bortz,Zbook}.    
Recently, it is studied by functional renormalization group \cite{FRG} and also experimentally in a quasi-one-dimensional conductor \cite{thermionic}.   
However, it is still rare that the finite-temperature thermodynamic behavior is explicitly and exactly shown by a theoretical method for a large but finite lattice system without making any approximation or assumption.

In this paper we study the finite-temperature behavior of an integrable model of the spin-1/2 anti-ferromagnetic XXZ periodic chain with a spin-1/2 impurity and make an additional report to Ref. \cite{YSD}.  In particular, we investigate how the crossover temperature $T_c$ for the impurity specific heat $c_{\rm imp}$ depends on the impurity parameter $x$ and show that the $x$-dependence of crossover temperature $T_c$ is consistent with the analytic expression derived by setting the impurity susceptibility $\chi_{\rm imp}$ to be proportional to the inverse of the crossover temperature: $1/T_c \propto \chi_{\rm imp}(x)$.  Here the impurity parameter $x$ is related to the coupling strength between the impurity spin and other interacting spins.     

In the XXZ impurity model we can exactly study the effect of an impurity embedded in the interacting quantum spins in one dimension, while in the Kondo model the effect of an impurity coupled with itinerant electrons in three dimensions.  The XXZ impurity model and its variants have been investigated by many authors.  Schlottmann studied the integrable spin-$S$ XXZ spin chain with one spin-$S'$ impurity \cite{Sch} in association with the multi-channel Kondo effect \cite{Schlottmann1991,Sacramento,Bortz}. 
The spin-1/2 impurity in the open spin-1/2 XXX chain was studied by Frahm and Zvyagin \cite{FZ1997} and it was shown that the Kondo-like temperature exists,  while the periodic XXZ chain with an impurity was studied by Eckle et al. \cite{Eckle}.    
The thermodynamic behavior of the spin-1/2 impurities in the spin-1/2 XXZ chain 
was shown in Ref. \cite{ZK2000} for several distributions of the impurity parameters. 
However, the analytic expressions of the impurity susceptibility and the impurity specific heat at low temperature in Ref. \cite{Sch} are not accurate enough to evaluate the Wilson ratio correctly, in particular,  when the impurity parameter is small.

The integrable model of the spin-1/2 XXZ periodic chain with a spin-1/2 impurity studied in Ref. \cite{YSD} corresponds to the case of  $S=S'=\fr12$ in the XXZ impurity model of Ref. \cite{Sch}.  The finite-temperature behavior of the integrable XXZ impurity model    is investigated by numerically solving the truncated integral equations of the thermal Bethe ansatz \cite{TS,T73,FZ,T}  and evaluate the specific heat and the entropy numerically \cite{YSD}. By plotting graphs of the impurity specific heat $c_{\rm imp}$ 
versus temperature $T$ and those of the impurity entropy versus temperature $T$,  
 it is shown that the impurity spin gradually becomes a free spin through pseudo-decoupling from other spins,  while in low temperature it couples strongly to them such as the Kondo effect  \cite{YSD}.  Here,  we call it {\it pseudo-decoupling},  if the exchange coupling between the impurity spin and other spins is very small but nonzero. 
It occurs in the present model when the absolute value of the impurity parameter is very large.    

It is shown that the Wilson ratio at low temperature is given by $r  ={2 \pi}/({\pi-\z})$ 
for the spin-1/2 anti-ferromagnetic XXZ periodic chain with a spin-1/2 impurity, where 
the XXZ anisotropy parameter $\Delta$ is given by $\Delta= \cos \z$ ( $0 \le \zeta < \pi$)  \cite{YSD}. 
It is independent of the impurity parameter $x$ if its absolute value is small enough with respect to the given temperature.  
Thus, the Hamiltonians of the XXZ impurity model with the same XXZ anisotropy parameter $\Delta$  
but different values of impurity parameter $x$ are classified in the same universality class \cite{YSD}. 

The contents of the paper consist of the following. In section 2 we derive the expression of the XXZ impurity Hamiltonian 
in terms of the local spin operators $S_j^a$. We show that the ground-state energy  evaluated by diagonalizing the 
XXZ impurity Hamiltonian expressed in terms of the local spin operators is consistent 
with that obtained by solving the Bethe ansatz equations numerically 
for several different values of the impurity parameter.  In section 3 we review the analytic expression of the impurity susceptibility 
at zero temperature derived by the Wiener-Hopf method in Ref. \cite{YSD}.  
In section 4, after we  briefly review the method for evaluating the specific heat via the thermal Bethe ansatz in Ref. \cite{YSD},  we plot the graphs of the impurity specific heat versus impurity parameter $x$ for five different values of temperature $T$. We then show numerically that the crossover temperature $T_c$ as a function of impurity parameter $x$ is derived by setting the inverse of $T_c$ to be equal to impurity susceptibility $\chi_{\rm imp}(x)$: $A/T_c = \chi_{\rm imp}(x)$ with some constant $A$. Finally, in section 5 we give concluding remarks.

\section{Integrable model  with an impurity}

\subsection{The XXZ Hamiltonian with an impurity}

We now formulate the integrable spin-1/2 anti-ferromagnetic XXZ chain with a spin-1/2 impurity 
through the algebraic Bethe ansatz. 
Let us denote by $V_0$, $V_1, \ldots, V_N$ the two-dimensional vector spaces over ${\bf C}$. 
For the $N$-site chain we consider the quantum space $V_1 \otimes V_2 \otimes \cdots \otimes V_N$, 
where the spin operators on the $j$th site  act on $V_j$ for $j=1, 2, \ldots, N$.  We call  $V_0$ the auxiliary space. 
We denote the transfer matrix of the XXZ spin chain by 
$\tau_{1\cdots N}(\lam|\xi_1,\cdots,\xi_N)$ acting on the quantum space 
with the spectral parameter  $\lambda$  and inhomogeneity parameters 
$\xi_1, \xi_2, \ldots, \xi_N$. We shall define it  shortly.  
The XXZ Hamiltonian with an impurity is given by the logarithmic derivative of an inhomogeneous  transfer matrix 
\begin{eqnarray}
\label{h}
 \begin{split}
\xxz(x)&=\left. {\frac{\sinh(\i \zeta)} 2} \dlam \log \tau_{1\cdots N}\left(\lam\Big|\frac{\i \zeta}{2}-x,
\frac{\i \zeta}{2},\cdots,\frac{\i \zeta}{2}\right)\right|_{\lam\to \frac{\i \zeta}{2}} 
 \end{split},
\end{eqnarray}
where we call the parameter $x$ the impurity parameter and we assume that it is real. 
The impurity Hamiltonian $\xxz(x)$ in \re{h} becomes the standard XXZ Hamiltonian when $x=0$.
We define the inhomogeneous transfer matrix $\tau_{1\cdots N}$ by the following product of the $R$-matrices 
\begin{equation}
 \tau_{1\cdots N}(\lam|\xi_1,\cdots,\xi_N)
 = \mbox{tr}_0 R_{0N}(\lam-\xi_N) \cdots R_{01}(\lam-\xi_1)
\end{equation}
where inhomogeneity parameters $\xi_1, \xi_2, \ldots, \xi_N$ are arbitrary 
and the $R$-matrices of the XXZ model acting on the tensor product $V_0 \otimes V_n$ are given by 
\begin{equation}
  R_{0n}(\lam)=
\begin{pmatrix}
1&&& \\
&b(\lam)&c(\lam)& \\
&c(\lam)&b(\lam)& \\
&&&1
\end{pmatrix}_{\!\![0,n]} \, . 
\end{equation}
Here, suffix $[0,n]$ means that the four-by-four matrix acts on  the tensor product space $V_0 \otimes V_n$
and functions $b(\lam)$ and $c(\lam)$ are expressed as   
\begin{eqnarray}
 b(\lam)=\frac{\sinh(\lam)}{\sinh(\lam+\i \zeta)}, \quad  
c(\lam)=\frac{\sinh(\i \zeta)}{\sinh(\lam+\i \zeta)} ,  
\end{eqnarray}
and parameter $\z$ is related to the XXZ anisotropy parameter $\Delta$ by $\Delta=\cos \zeta$ with $0 \le \zeta < \pi$.

Through straightforward calculation we obtain the following expression of the impurity Hamiltonian 
$\xxz(x)$ in terms of local spin operators $S_j^{\pm}$ and $S^z_j$ on the $j$th sites \cite{YSD}. 
\begin{eqnarray}\label{eq:hxxz} 
 \xxz(x)
&=&\sum_{n=2}^{N-1}  \left[ \frac12(S_n^+S_{n+1}^-+S_n^-S_{n+1}^+)
+\Delta\left(S_n^zS_{n+1}^z-\frac14\right)\right] \nn \\
&+&c^+c^-
\left[ {\frac{\cosh x}2} (S_N^+S_{1}^-+S_N^-S_{1}^+)
+\Delta\left(S_N^zS_{1}^z-\frac14\right)\right] \nn \\
&+&c^+c^-\left[
{\frac{\cosh x}2} (S_1^+S_2^-+S_1^-S_2^+)+\Delta S_1^zS_2^z\right] \nn \\
&+&b^+b^-\left[
\frac{\D}2(S_N^+S_2^-+S_N^-S_2^+)+\Delta S_N^zS_2^z\right] 
-\fr{\Delta}{4} \nn \\
&+&b^+c^-\Biggl[
 \cosh x   (S_N^+S_2^--S_N^-S_2^+)S_1^z \nn \\
 &&\ \ \ \ \ \ \ \ -\D(S_N^+S_1^--S_N^-S_1^+)S_2^z \nn \\
 &&\ \ \ \ \ \ \ \ -\D(S_1^+S_2^--S_1^-S_2^+)S_N^z \Biggr], 
\end{eqnarray}
where  symbols $b^\pm$ and $c^\pm$ are given by  
\begin{eqnarray}\label{bc}
 b^\pm=b(\pm x)=\frac{\sinh x}{\sinh( x \pm i\zeta)},
 \ \ \ \ 
 c^\pm=c(\pm x)=\pm\frac{\sinh(i\zeta)}{\sinh(x\pm i\zeta)} \, ,  
\end{eqnarray}
and the spin operators $S_j^{\pm}$ and $S_j^z$ are expressed in terms of the 
Pauli matrices  as  
\begin{equation} 
S_j^{\pm} = {\frac 1 2} \sigma_j^x \pm i {\frac 1 2} \sigma_j^y , \quad 
 S_j^z = {\frac 1 2 } \sigma_j^z \, ,  \quad  \mbox{for} \quad j = 1, 2, \ldots, N \, .  
\end{equation}  

We confirm in the explicit expression \re{eq:hxxz}  
that the impurity Hamiltonian  $\xxz(x)$  gives the spin-1/2 anti-ferromagnetic XXZ chain 
if  impurity parameter $x$ vanishes.  It also follows from \re{eq:hxxz}  that 
the impurity Hamiltonian $\xxz(x)$ is Hermitian for any real value of impurity parameter $x$.

In the limit of sending the impurity parameter to infinity: $x\to\infty$, 
the impurity Hamiltonian \re{eq:hxxz} reduces to the following:  
\begin{eqnarray}\label{hinf}
 \xxz(\infty)
 &=&\sm{n=2}{N-1}  \Bigl[ \frac12(S_n^+S_{n+1}^-+S_n^-S_{n+1}^+)
 +\D\(S_n^zS_{n+1}^z-\frac14\)\Bigr] \nn \\
 && +\frac\D2\(S^+_NS^-_2+S^-_NS^+_2\)+\D\(S_N^zS_2^z-\frac14\) \nn \\
 && +2\sqrt{1-\D^2}\(S^x_NS^y_2-S^y_NS^x_2\)S_1^z. 
\end{eqnarray}
Furthermore, in the XXX limit: $\Delta \to 1$ it becomes 
\begin{eqnarray}\label{xxxinf}
H_{\rm XXX}(\infty) = 
 \sm{n=2}{N-1} \left[ S_n^xS_{n+1}^x+S_n^yS_{n+1}^y+S_n^zS_{n+1}^z-\fr14\right]  +  \left( S_N^x S_{2}^x+S_N^yS_{2}^y+S_N^zS_{2}^z-\fr14 \right). 
\end{eqnarray}
Thus, the impurity spin operators $S_1^a$ ($a=x,y,z$)  are decoupled from those of other sites.


\subsection{The Bethe ansatz equations with an impurity}

Let us define functions $\theta_n(z)$ for positive integers $n= 1, 2, \ldots, $ by 
\begin{equation}
 \theta_n(z) =  \i \log \left[-\frac{\sinh(z+\frac{\i n \z}{2})}{\sinh(z-\frac{\i n\z}{2})}\right] \, . 
\end{equation}
The Bethe ansatz equations (BAE) for the impurity Hamiltonian (\ref{h}) 
in the $M$ down-spin sector are given by 
\begin{equation}
(N-1) \theta_1(z_l) + \theta_1(z_l+x) =   
2\pi I_l +\sum^M_{j=1; j\neq l}\theta_2(z_l-z_j) ,  \quad 
\mbox{for} \, \,   \l=1,\ldots, M . \label{bae}  
\end{equation}
Here $I_l$ are called the Bethe quantum numbers, and they are given by 
integers or half-integers according to the following rule:  
\begin{equation}
 I_l \equiv \frac{N-M+1}{2} \pmod{1}, \ \ \mbox{for} \, \,  l=1, 2, \ldots, M . 
\end{equation}
We note that functions $\theta_n(z)$ are expressed in terms of the arctangent function as follows. 
\begin{equation} 
\theta_n(z) = 2 \tan^{-1} \left(  \cot \left( \frac {n \z} 2 \right)   \tanh z \right).   
\end{equation}

By solving BAE (\ref{bae}) numerically we can evaluate the energy eigenvalues of the Bethe ansatz eigenvectors with $M$ down-spins. 
We call a solution of BAE  $\{ z_{\ell}| \, \ell=1, 2, \ldots, M \}$ the Bethe roots.  
The energy $E$ of a Bethe ansatz eigenstate with the Bethe roots $\{ z_{\ell} \}$ is expressed as  
\begin{eqnarray}\label{en}
 E= -\ds{\sum^M_{l=1}\fr{ \sin^2 \z}{\cosh 2z_l-\cos\z}}\, .  
 \end{eqnarray}
We note that the energy depends on impurity parameter $x$,  
since the solution $\{ z_j \}$ depends on  the parameter $x$ through BAE in (\ref{bae}).

\subsection{Ground state of the XXZ spin chain with an impurity}

It is not clear how the Bethe quantum numbers $\{ I_j \}$ should depend on the impurity parameter $x$ for a given 
Bethe ansatz eigenstate of the XXZ impurity Hamiltonian. Here we recall that the Bethe quantum numbers appear in BAE \re{bae}. 
It is therefore not trivial whether the Bethe quantum numbers  for the ground state with $x=0$ 
also gives the Bethe quantum numbers for the ground state with $x \ne 0$. 
%

We have a conjecture that the Bethe quantum numbers of the ground state do not change if impurity parameter $x$ is real 
even when  $x \ne 0$.  By taking advantage of the explicit expression \re{eq:hxxz} of the impurity Hamiltonian 
we can evaluate its eigenvalues through exact numerical diagonalization, i.e. we can numerically diagonalize the impurity Hamiltonian through the expression \re{eq:hxxz}.   We thus confirm that the set of the Bethe quantum numbers of the ground state for $x=0$ 
gives those in the case of $x \ne 0$. 

\begin{table}[h]
\begin{center}
  \scalebox{0.80}{
 \begin{tabular}{|c||c|c|c|c|c|c|} \hline
  $x$ & 0 & 1 & 2 & 3 & 4 & 5 \\ \hline
\hline 
  exact diagonalization & -4.392695167 & -3.961714971 &
  -3.743119726 & -3.680990863 & -3.660728144 & -3.653612539 \\ \hline 
  Bethe ansatz & -4.392695167 & -3.961714971 &
  -3.743119726 & -3.680990863 & -3.660728144 & -3.653612539 \\ \hline
 \end{tabular}}
 \caption{Ground-state energy of the XXZ  impurity Hamiltonian \re{eq:hxxz} with $N=8$ and $\D=0.6$ for six values of impurity parameter ($x=0, 1,\cd,5$) evaluated by exact numerical diagonalization and by the Bethe ansatz. Here we assume that the Bethe quantum numbers of the ground state do not depend on impurity parameter $x$ and are given 
by $\{\pm 1/2, \pm 3/2\}$.  }
 \label{table}
\end{center}
\end{table}

We have thus confirmed that the lowest energy level of \re{eq:hxxz} is completely 
consistent with the eigenvalue evaluated from BAE \re{bae} in the cases of $N \leqq 10$ with respect to numerical errors.  For an illustration, some results for $N=8$ are shown up to 10 decimal degits in Table \ref{table}. They are obtained by assuming the same set of the Bethe quantum numbers.  We have another conjecture  that this correspondence is valid for any larger number of sites.

\section{Impurity susceptibility through the Wiener-Hopf method}

We now derive the analytic expression of the impurity magnetic susceptibility at zero temperature. 
We add a small magnetic field $h$ to the impurity Hamiltonian  \re{eq:hxxz}:
\begin{eqnarray}\label{hh}
H^{'}=\xxz(x)-2h\sum_{n=1}^N S_n^z.
\end{eqnarray}
By taking the thermodynamic limit of BAE \re{bae} through the Euler-Maclaurin formula,  
we derive the following integral equation: 
\begin{eqnarray}\label{baeh}
 \rho(z, x) + \int^B_{-B}a_2(z-z')\rho(z', x)dz'
 =\frac{1}{N} \lt (N-1)a_1(z)+a_1(z+x) \rt.
\end{eqnarray}
Here parameter $B$ denotes the Fermi point determined by the magnetic field $h$, 
and $\rho(z, x)$  the density of the Bethe roots in the ground-state solution 
of BAE (\ref{bae}), 
where functions $a_n(z)$ for $n=1, 2, \ldots, $ denote the derivatives of functions $\theta_n(z)$:  
$a_n(z)={\theta_n^{'}(z)}/{2\pi}$ and they given by 
\begin{eqnarray}
 a_n(z) 
=-\frac{\i}{2\pi}
 \frac{\sinh(\i n \z)}{\sinh(z+\frac{\i n \z}{2})\sinh(z-\frac{\i n \z}{2})}. \label{eq:an}
\end{eqnarray}
We note that  we have the infinite Fermi point, $B=\infty$, for $h=0$.  

The magnetization per site $s^z(x)$ and the energy per site $e(x)$ 
are written in terms of the integrals of the root density $\rho(z,x)$ as follows. 
\begin{eqnarray}
 s^z(x) &=&\fr12-\int_{|z|<B} \rho(z,x) dz \nonumber \\ 
&  = & \fr{\pi}{2\pi-2\z}\int_{|z|>B} \rho(z,x) dz, \\
 e(x) &=& \pi \sin \z \int_{|z|>B}\sigma_0(z)\rho(z,x)dz,  
\end{eqnarray}
where  the bulk root density $\sig_0(z)$ is given by 
\begin{equation} 
 \sig_0(z)=\fr{1}{2\z}\mathrm{sech}\fr{\pi z}{\z}.
\end{equation}
Through the Wiener-Hopf method \cite{YSD} the magnetization per site $s^z(x)$ is given by   
\begin{eqnarray}
 s^z(x)
 &\simeq&\fr{\i \pi}{N\z(\pi-\z)}G_+(0) \fr{G_-\(-\fr{\i \pi}{2}\)}{\fr{\i \pi}{\z}} \fr{G_+(0) h\z}{(\pi-\z)\sin \z G_+\(\fr{\i \pi}{2}\)}
 \(N-1+\cosh \fr{\pi x}{\z}\) \nn \\ 
&=& \fr{2\z h}{N\pi(\pi-\z) \sin \z}\(N-1+\cosh\fr{\pi x}{\z}\) .
\label{eq:mps}
\end{eqnarray} 
Here, function $G_{+}(k)$ ($k= \i \pi z$) is expressed  in terms of the gamma function $\Gamma(z)$ by 
\begin{equation}
 G_+(\i \pi z)
 =\sqrt{2\pi \left(1-\fr{1}{\g} \right) } \( \fr{(\g-1)^{\g-1}}{\g^{\g}}\)^z
 \fr{\Gamma(\g z+1)}{\Gamma({\fr 1 2}+z)\Gamma((\g-1)z+1)}, 
\end{equation}  
and $\g$ is given by $\g={\pi}/{\z}$, and function $G_{-}(k)$ is determined by the following relation:  
\begin{equation} 
 G_-(k)=G_+(-k) \, . 
\end{equation}
In equation \re{eq:mps} we have assumed that magnetic field $h$ is very small, so that we consider only linear terms of $h$.  
The Fermi point $B$ is determined as a function of $h$ by the following condition: 
\begin{eqnarray}
 \fr{\partial}{\partial B} \left( e(x) - 2 h s^z(x) \right) = 0 .
\end{eqnarray}
We therefore have 
\begin{eqnarray}
 \exp\left( {-\fr{\pi B}{\z}} \right)
 =\fr{G_+(0) \z h}{(\pi-\z)\sin \z G_{+} \(\fr{\i \pi}{2}\)}.  \label{eq:h-B}
\end{eqnarray}

Let us define the impurity magnetization  $s_{\rm imp}^z(x)$ at zero temperature by 
\begin{equation} 
 N s^z(x) = (N-1) s^z(0) + s_{\rm imp}^z(x) \, . 
\end{equation}
Here,  $N s^z(x)$ gives the total magnetization of the impurity XXZ spin chain 
with impurity parameter $x$.  We also express it by $S_{\rm tot}^z$.  
We then define the impurity susceptibility $\chi_{\rm imp}(x)$ by 
\begin{equation} 
\chi_{\rm imp} = \left. \frac {\partial \, 2s^{z}_{\rm imp}(x)} {\partial h} \right|_{h=0} .   
\end{equation} 
The impurity susceptibility at zero temperature is therefore given by 
\begin{equation} 
\label{eq:sc-imp}
\chi_{\rm imp}(x)= \fr{4 \z }{\pi(\pi-\z) \sin \z} \cosh\fr{\pi x}{\z} .
\end{equation}

\section{Impurity specific heat}

\subsection{Coupled integral equations of the thermal Bethe ansatz}

We now evaluate the crossover temperature by numerically evaluating   the specific heat per site $c(x,T)$ of the impurity XXZ spin chain at temperature $T$. We first define the impurity specific heat  $c_{\rm imp}(x, T)$ at temperature $T$ with impurity parameter $x$ by \cite{YSD} 
\begin{equation} 
 N c(x,T) = (N-1) c(0, T) + c_{\rm imp}(x, T) \, . 
\end{equation}
Here,  $N c(x, T)$ gives the total specific heat of the impurity XXZ spin chain 
with impurity parameter $x$ at temperature $T$.

Let us consider the string solutions for the BAE at $\z={\pi}/{3}$. According to the string hypothesis \cite{TS,T},  
there are only three types of solutions: 1-string solutions with even parity (i.e., real solutions) $\{ z_j^{(1+)} \}$, 
2-string solutions  $\{ z_j^{(2)} \pm \i \z/2 \}$, and real solutions of  1-string with odd parity  $z_j^{(1 -)} + \i \pi/2$.  
Here we assume that all the three types of string centers, $z_j^{(1+)}, z_j^{(2)}$ and $z^{(1-)}$ are given by real numbers. 
 
For the density functions of the three types of string solutions appearing at $\z=\pi/3$ \cite{TS}
a set of coupled integral equations truncated at $\z=\pi/3$ 
are derived through the thermal Bethe ansatz, which will be formulated shortly. 
We express the free energy in terms of the solution to the thermal Bethe ansatz equations. 
By taking the second derivative of the free energy with respect to temperature 
we derive the integral expression for the specific heat of the model \cite{FZ}. 
Thus, by solving the thermal Bethe ansatz equations numerically, 
we obtain the specific heat of the impurity XXZ spin chain denoted by $c(x,T)$ as follows. 
\begin{eqnarray}
 c(x, T)=\fr{1}{N} \intall \lt \a_1 v_1
 +\a_2 \(2v_2+b_1*v_1+2b_2*v_2 \) \rt dz, \label{imp:sh}
\end{eqnarray}
where  symbol  $*$ denotes the convolution and functions $\alpha_j(z)$ are given by 
\begin{equation}
 \a_j(z) = (N-1)a_j(z)+a_j(z+x) \quad \mbox{for} \, \, j=1, 2.   
\end{equation} 
Here functions  $a_j(z)$ and $b_j(z)$ are given by 
\begin{eqnarray}
& a_j(z)=\displaystyle{\fr{1}{2\pi}\fr{2\sin \z q_j}{\ch 2z + \cos \z q_j}} \, , 
 \hspace{5mm} q_1=2, \ \ q_2=1 , \\
 & b_1(z) = \displaystyle{ \fr{3 \ch \left( {3z}/{2} \right) }{\sqrt{2} \pi \ch 3z}}, \hspace{10mm}  
\displaystyle{b_2(z)=\fr{3}{4\pi \ch \left( {3z}/{2} \right) }} .
\end{eqnarray} 
With $\k = -\pi \sin \z$, functions $v_j$ in \re{imp:sh} 
are given by the solutions of the following equations. 
\begin{eqnarray}
 &&\left\{
 \begin{array}{lll}
 v_1=\fr{1}{T^2}\fr{{u_1}^2}{e^{h_1}-1}
 +\(1-e^{-h_1}\) \[-2b_2*v_1-2b_1*v_2  \] \\
 v_2=\fr{1}{T^2}\fr{{u_2}^2}{e^{h_2}-1}
 +\(1-e^{-h_2}\) \[-b_1*v_1-2b_2*v_2   \] 
 \end{array}
 \right. \, , \label{int:v} \\
 &&\left\{
 \begin{array}{lll}
 u_1=\(1-e^{-h_1}\) \[-\k b_1-2b_2*u_1-2b_1*u_2  \] \\
 u_2=\(1-e^{-h_2}\) \[-\k b_2-b_1*u_1-2b_2*u_2  \]
 \end{array}
 \right., \label{int:u}  \\
 &&\left\{
 \begin{array}{lll}
 h_1=\log \[ 1+\exp \[-\fr{\k}{T}b_1-2b_2*h_1-2b_1*h_2  \] \] \\
 h_2=\log \[ 1+\exp \[-\fr{\k}{T}b_2-b_1*h_1-2b_2*h_2  \] \]
 \end{array}
 \right. .  \label{int:h}
\end{eqnarray}

We can solve the thermal Bethe ansatz equations (\ref{int:v}), (\ref{int:u}) and (\ref{int:h}) as follows. Firstly, we solve the coupled integral equations \re{int:h} for $h_1$ and $h_2$. Secondly, we solve equations \re{int:u} for $u_1$ and $u_2$ from the result of  $h_1$ and $h_2$. Thirdly, 
we solve equations \re{int:v} for $v_1$ and $v_2$, from which we have the 
estimates of specific heat $c(x,T)$ and then $c_{\rm imp}(x, T)$.    

\subsection{Crossover temperature for impurity specific heat versus impurity parameter}

\begin{figure}[htb]
 \centering
 \includegraphics[width=10cm]{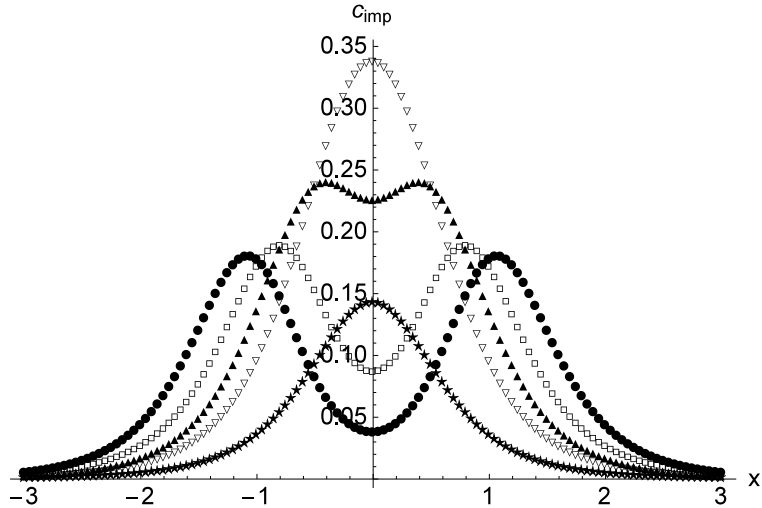}
 \caption{Impurity specific heat $c_{\rm imp}$ versus impurity parameter $x$ at five different temperatures $T$. The data points for $T=10^{-4/3}, 10^{-1}, 10^{-2/3}, 10^{-1/3}$ and $10^{0}$ are depicted  by filled circles, open squares, filled upper triangles, open lower triangles and filled stars, 
respectively.  
}
 \label{c_imp}
\end{figure}

In Figure \ref{c_imp}  the estimates of impurity specific heat  $c_{\rm imp}(x, T)$ are plotted against impurity parameter $x$ 
 for five fixed values of temperature $T$: $T=10^{-4/3}, 10^{-1}, 10^{-2/3}, 10^{-1/3}$ and $10^{0}$. 

Two peaks are clearly seen in the graph of $c_{\rm imp}$ versus $x$ 
at the lowest temperature $T=10^{-4/3}$ (depicted by filled circles) in Figure \ref{c_imp}.   
The positions of the two peaks are symmetric with respect to the origin of the $x$-axis ($x=0$):   
$x = x_{\rm peak}(T)$ and $x = - x_{\rm peak}(T)$. They are approximately given by $\pm x_{\rm peak}(T) = \pm 1.1$.  It is clear in Figure \ref{c_imp} that as temperature $T$ increases the two peak positions  $\pm x_{\rm peak}(T)$ become closer to the origin: $x=0$, while the peak height becomes taller.  The peak height increases until the two peak positions   $\pm x_{\rm peak}(T)$  merge to the origin ($x=0$). After the peak positions  $\pm x_{\rm peak}(T)$  reach at the origin ($x=0$), the peak height decreases rapidly.  The peak height of $c_{\rm imp}$ at $x=0$ decreases from 0.34 to 0.15 when temperature $T$ increases from $10^{-1/3}$ to $10^0$, as shown in Figure \ref{c_imp}.

Over the most of the region between the two peaks at $x=\pm x_{\rm peak}$ in the graph of impurity specific heat  $c_{\rm imp}$ versus  impurity parameter $x$ at temperature $T$, as shown in Figure \ref{c_imp}, we assume that the estimates of $c_{\rm imp}$ as a function of $x$ are consistent with the following expression  
\begin{equation} 
\label{eq:c-imp}
c_{\rm imp}(x, T)= {\fr{2 \z T}{3 \sin \z}} \cosh \fr{\pi x}{\z} \, .  
\end{equation}
Here we recall that the consistency has already been shown in Figure 6 of Ref. \cite{YSD}. 
Thus, at a given temperature $T$ impurity specific heat $c_{\rm imp}(x, T)$ as a function of  impurity parameter $x$ is consistent with that of the low-temperature regime if the absolute value of $x$ is smaller than the peak position $x_{\rm peak}(T)$ of the graph of $c_{\rm imp}$ versus $x$ at the given temperature $T$, i.e. if  $|x| < x_{\rm peak}(T)$. If the absolute value of impurity parameter $x$ is larger than the peak position at temperature $T$, i.e. if  $|x| > x_{\rm peak}(T)$, we expect that the coupling between the impurity spin and other spins is too weak with respect to thermal fluctuations at temperature $T$, so that impurity specific heat  $c_{\rm imp}$ should almost vanish. In Figure \ref{c_imp}  impurity specific heat  $c_{\rm imp}$ indeed becomes small exponentially with respect to  $x$ in the region where we have  $|x| > x_{\rm peak}(T)$.    

Based on the above observations, we now define the crossover temperature $T_c$ (or $T_c(x)$)  
as a function of  impurity parameter $x$ for impurity specific heat $c_{\rm imp}$. 
For a given value of impurity parameter $x$ we define  the crossover temperature $T_c(x)$ 
by a temperature $T$ such that the peak position at the temperature $T$ 
is equal to the given value of the impurity parameter: $x_{\rm peak}(T)= x$.  It thus follows that 
we have  $x_{\rm peak}(T_c(x))= x$.

\begin{figure}[htb]
 \centering
 \includegraphics[width=8cm]{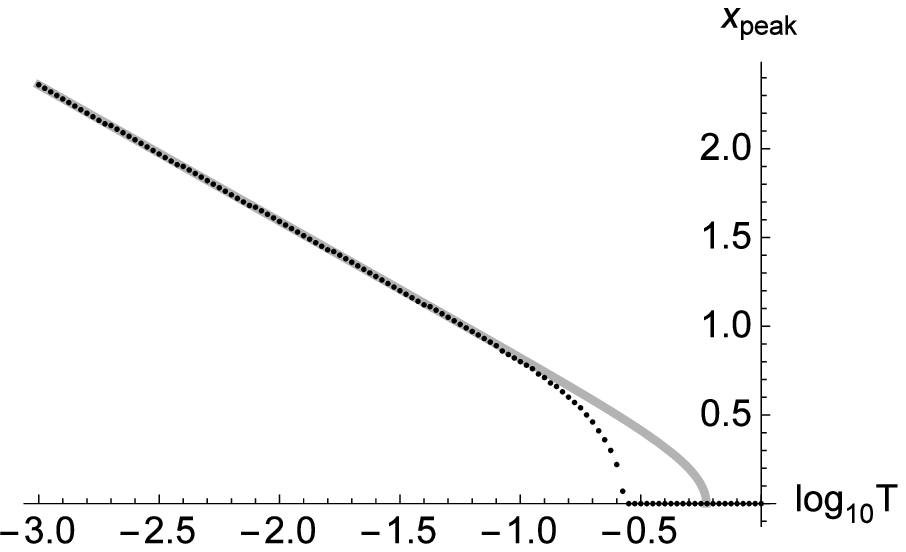}
 \caption{$\vtop{
\hbox{Peak position $x_{\rm peak}(T)$ versus $\log_{10} T$ for temperature $T$. } 
\hbox{The thick gray line is given by eq. \re{eq:direct} with $\z = \pi/3$ and $A=0.435$. }
}$  } 
 \label{logT}
\end{figure}

In Figure \ref{logT} the estimates of peak position $x_{\rm peak}(T)$ 
in the graph of $c_{\rm imp}$ versus $x$ at temperature $T$ are plotted against  $\log_{10} T$. 
The linear behavior of the data plots in Fig. \ref{logT} with lower temperatures  
is consistent with  the inverse of  the analytic expression \re{eq:sc-imp} of impurity susceptibility $\chi_{\rm imp}(x)$. 
By setting  $x=x_{\rm peak}(T_c)$ and $\chi_{\rm imp}(x)= A/T_c$ with some constant $A$ we have 
\begin{equation}
\frac A {T_c} = \fr{4 \z }{\pi(\pi-\z) \sin \z} \cosh\fr{\pi x_{\rm peak} }{\z} .
\end{equation} 
We express $x_{\rm peak}$ as a function of $T_c$ as   
\begin{equation}
x_{\rm peak} = \frac {\z} {\pi} \cosh^{-1}\left( A  \frac {\pi (\pi-\z) \sin \z}  {4 \z T_c} \right) \, . \label{eq:direct}
\end{equation}
Here $A$ denotes a constant to shift the graph of  $x_{\rm peak}$ versus $\log T$. 
As shown in Figure \ref{logT}, the graph of $x_{\rm peak}$ versus $\log_{10} T$ 
is consistent with the graph  of  \re{eq:direct} in lower temperatures such as 
for $\log_{10} T < -1.0$ .

When $x \gg 1$ we can approximate $\cosh \left( {\pi x}/{\z} \right)$ by $\exp\left( {\pi x}/{\z} \right)/2$.  
By expressing  $x_{\rm peak}$ as a function of $T_c$ we have  
\begin{equation}
 x_{\rm peak} \approx - \frac {\z} {\pi} \ln T_c +  \frac {\z} {\pi} \ln \left(A \fr{\pi(\pi-\z) \sin \z}{2 \z } \right). 
\end{equation}  
It is consistent with the linear part of the graph of $x_{\rm peak}$ versus $\log_{10} T$ 
for lower temperatures such as $\log_{10} T < -1.0$,  as shown in  Figure \ref{logT}.

\section{Concluding remarks}

We have investigated how  the crossover temperature $T_c$ for the impurity specific heat $c_{\rm imp}(x, T)$ depends on the impurity parameter $x$. We have determined $T_c$ through the two peak positions $\pm x_{\rm peak}(T)$ in the graph of  $c_{\rm imp}(x, T)$ versus $x$ at temperature $T$: For a given value of $x$ we define 
$T_c$ by the condition: $x_{\rm peak}(T_c) =x$. 
We have shown that the $x$-dependence of crossover temperature $T_c$ 
is consistent with the analytic expression derived by setting  the inverse of $T_c$ to be proportional to impurity susceptibility $\chi_{\rm imp}(x)$: $1/{T_c} \propto \chi_{\rm imp}(x)$.   

\ack
The present study is partially supported by Grant-in-Aid for Scientific Research 
No. 15K05204.

\end{document}